\documentclass[sigconf]{acmart}
\AtBeginDocument{%
  \providecommand\BibTeX{{%
    \normalfont B\kern-0.5em{\scshape i\kern-0.25em b}\kern-0.8em\TeX}}}

\setcopyright{acmlicensed}
\copyrightyear{2024}
\acmYear{2024}
\acmDOI{XXXXXXX.XXXXXXX}
\acmConference[DigiPro '24]{The Digital Production Symposium}{27 July 2024}{Denver, CO}

\begin{document}

\title{Implementing a Machine Learning Deformer for CG Crowds: Our Journey}


\author{Bastien Arcelin}
\authornote{Both authors contributed equally to this research.}
\affiliation{%
  \institution{Golaem}
  \streetaddress{31 rue de la Frébardière}
  \city{Chantepie}
  \country{France}}
  \postcode{35135}
\email{bastien.arcelin@golaem.com}
\orcid{0000-0002-3553-792X}

\author{Sebastien Maraux}
\authornotemark[1]
\affiliation{%
  \institution{Golaem}
  \streetaddress{31 rue de la Frébardière}
  \city{Chantepie}
  \country{France}}
  \email{sebastien.maraux@golaem.com}
\orcid{0009-0007-5143-3120}

\author{Nicolas Chaverou}
\orcid{0009-0003-8659-0880}
\affiliation{%
  \institution{Golaem}
  \streetaddress{31 rue de la Frébardière}
  \city{Nouméa}
  \country{New Caledonia}}
\email{nicolas.chaverou@golaem.com}

\renewcommand{\shortauthors}{Arcelin et al.}

\copyrightyear{2024}
\acmYear{2024}
\setcopyright{acmlicensed}\acmConference[DigiPro '24]{The Digital Production Symposium}{July 27, 2024}{Denver, CO, USA}
\acmBooktitle{The Digital Production Symposium (DigiPro '24), July 27, 2024, Denver, CO, USA}
\acmDOI{10.1145/3665320.3670994}
\acmISBN{979-8-4007-0690-5/24/07}

\begin{abstract}
  CG crowds have become increasingly popular this last decade in the VFX and animation industry: formerly reserved to only a few  high end studios and blockbusters, they are now widely used in TV shows or commercials. Yet, there is still one major limitation: in order to be ingested properly in crowd software, studio rigs have to comply with specific prerequisites, especially in terms of deformations. Usually only skinning, blend shapes and geometry caches are supported preventing close-up shots with facial performances on crowd characters. We envisioned two approaches to tackle this: either reverse engineer the hundreds of deformer nodes available in the major DCCs/plugins and incorporate them in our crowd package, or surf the machine learning wave to compress the deformations of a rig using a neural network architecture. Considering we could not commit 5+ man/years of development into this problem, and that we were excited to dip our toes in the machine learning pool, we went for the latter. 

  From our first tests to a minimum viable product, we went through hopes and disappointments: we hit multiple pitfalls, took false shortcuts and dead ends before reaching our destination. With this paper, we hope to provide a valuable feedback by sharing the lessons we learnt from this experience.

\end{abstract}



\begin{teaserfigure}
 \centering
   \includegraphics[trim = {0cm 0cm 0cm 0cm}, clip, width=0.8\textwidth]{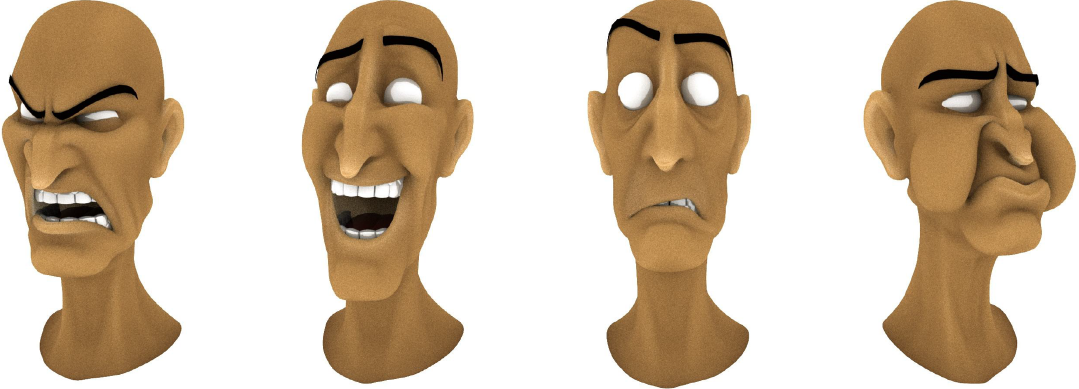}
       \caption{\centering{Examples of facial deformations using our Machine Learning Deformer in \textit{Golaem Crowd} © Stim Studio }}
    \label{fig:edgar_faces}
\end{teaserfigure}

\maketitle
\section{Context}
Founded in 2009 and based in France, \textit{Golaem} is an innovative software company that grew out of the \textit{INRIA} research center. It develops technologies for crowd simulation and character layout for the VFX and animation industry. 

\section{Why Do We Need A ML Deformer For CG Crowds?}
With the democratization of CG crowds, studios are now expecting crowd characters to blend seamlessly with keyframed hero characters, especially in terms of facial performances. However, they always hit the same limitation: only a subset of deformers are supported in their crowd package - usually skinning, blend shapes and geometry caches. Even if they sample every single rig deformation as a blend shape, those never reach the quality of their hero rig, it generates inaccurate deformations when blended or interpolated and limits edition when using rig inversion.

Fortunately, \citet{dreamworks_body} paved the way for a new avenue: approximating character deformations using Neural Networks (NN). This approach presents two main benefits: it allows to increase the rig performances (x5-x10) and to make it portable to other applications. While we were firstly interested in the portability aspect, we were also seduced by the speed up capabilities it promised. What if we could kill two birds with one stone? i.e. use the same technology to bring hero quality deformations to our crowd package and propose a new tool for animators to run their rigs faster while preserving the deformations.
 
In this paper, we start by presenting our prototype based on \cite{dreamworks_body}, describe the different challenges we faced and why we had to focus on another approach \cite{bluesky}. Then, we discuss the challenges of productizing a NN based application. The implementation of the method also raised some questions on the speed increase when applied to production rigs. Finally, we describe our implementation in our crowd package. 

\section{Our First ML Deformer}
\subsection{The Promise}
\citet{dreamworks_body} introduced a new method to approximate non-linear deformations. It uses Dense Neural Networks (DNN) and Principal Component Analysis (PCA) to approximate rig deformations. Authors propose to divide the global deformation into two parts: a linear and a non-linear one. The non-linear deformation being the most computationally expensive while animating, they approximate it with a DNN. It is a ground breaking work, even if it is limited to deformations related to skeleton animation. A lot of quality results are displayed in the paper and authors advertise great speed performance gain. They show that evaluating the deformation goes up to 5 to 10 times faster with their method than with an optimized evaluation engine. It looked like a great first step toward more complex deformations in our crowd package.

\subsection{The Reality}
\begin{figure}
    \centering
    \includegraphics[trim = {0cm 0cm 0cm 0cm}, clip, width = \columnwidth]{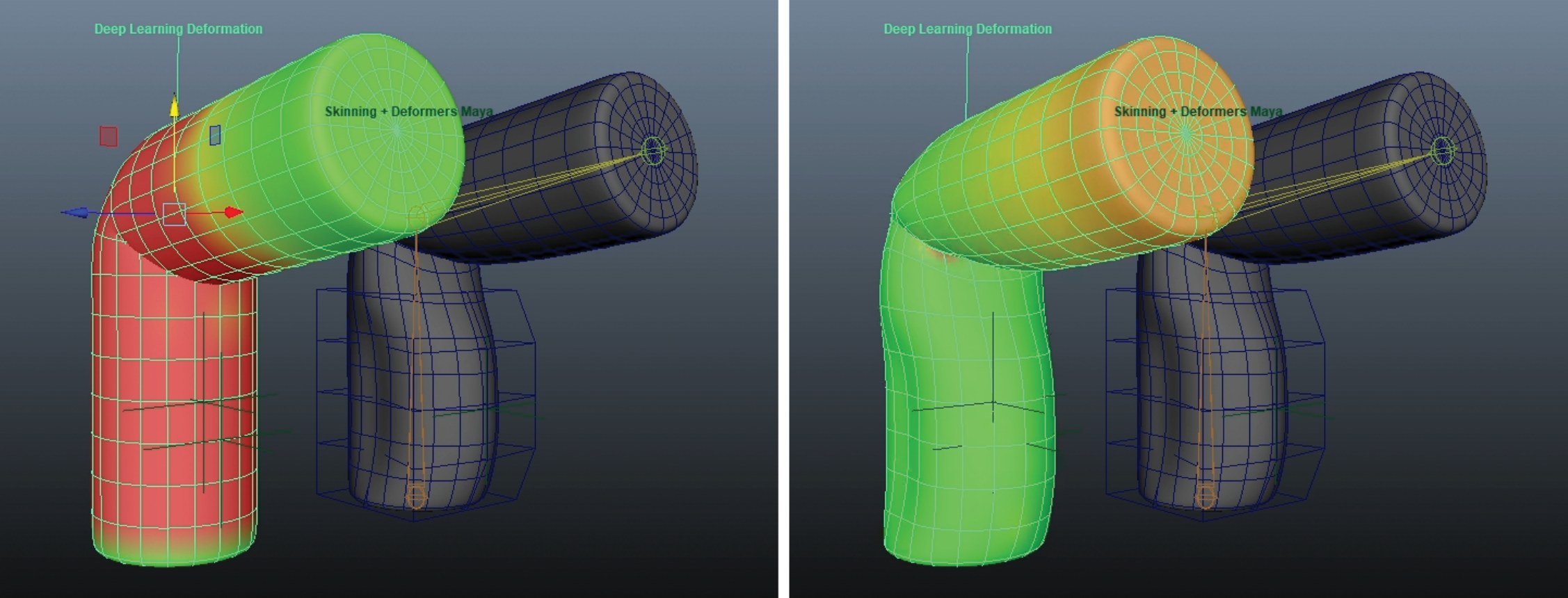}
    \caption{\centering{Example of an arm geometry with LBS (left) and with our ML deformer (right). Errors compared to the ground truth are represented with a heat map.}}
\label{fig:arm_twisting}
\end{figure}

Consequently, we implemented this method and checked if we were be able to replicate the results presented in the paper. We started with simple tests to validate our understanding of the method, such as a twisting and bulging arm (see Fig \ref{fig:arm_twisting}). 

\begin{figure}
    \centering
    \includegraphics[trim = {0cm 0cm 0cm 0cm}, clip, width = 1\columnwidth]{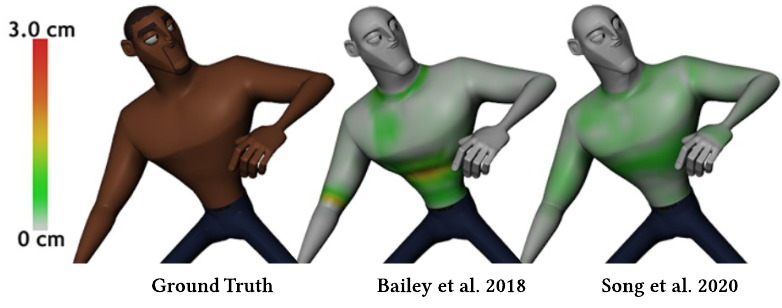}
\caption{\centering{Inference differences between \citet{dreamworks_body} and \citet{bluesky}. Errors compared to the ground truth are represented with a heat map. }}
\label{fig:comparaison_methods_deformation}
\end{figure}

While it worked as expected, we bumped into issues regarding how data are extracted for the dataset. For the input rotations, the paper uses Euler angles. Those being discontinuous, it lead to poor deformation results on edge cases and bad generalization. This may be good enough in a studio environment where riggers can adapt their rigs to this constraint but not for the off-the-shelf product we're targeting. Quaternions have the same limitation: one rotation can be represented with different values and mislead the training. As NN better fit continuous representations, we tackled this by storing rotations in a 6D representation composed of two vectors (front and up) \cite{Zhou_2019_CVPR}. Regarding the output geometry, vertices are represented in the local axis of the bone with the highest skinning influence. This lead to deformation errors between zones bounded to different bones (see Fig.\ref{fig:comparaison_methods_deformation}); it is especially noticeable for human eye as it is not a progressive error that spread over a zone, but a  localized error. 

Finally, this method has one major drawback: it is not accurate enough for facial animation. As a character head is usually bound to a small set of bones but produces a really large palette of deformations, the model can not fit those complex deformations. Thus, at the beginning of 2022, we concluded that this method was not advanced enough for our requirements and started to look for new works and luckily for us, we had several options.

\section{Back to the State of the Art, what Else Can We Implement?}

At SIGGRAPH 2020, several papers proposed new methods to approximate non-linear facial deformations. Following their previous work, the same team developed a new method specifically designed for facial animation \cite{dreamworks_face}. It is based on Convolutional Neural Networks (CNN) which infer deformation maps from the rig parameters. These maps are used to compute vertex offsets from the neutral pose applying bilinear interpolation at each vertex position in the UV coordinate space. On the other hand, \citet{bluesky} uses a simple NN architecture, but handles both facial and body deformations. It uses differential coordinates, a mesh representation that is obtained using the Laplacian operator \cite{sorkine_laplacian}, to learn the complete mesh deformation from rig controllers, and refine it with other smaller specialized networks. Finally, \citet{facebaker} proposed a method to approximate facial blendshapes from rig controller values using dense NN and PCA.

\begin{figure*}
    \centering
    \includegraphics[trim = {0cm 0cm 0cm 0cm}, clip, width = 0.75\textwidth]{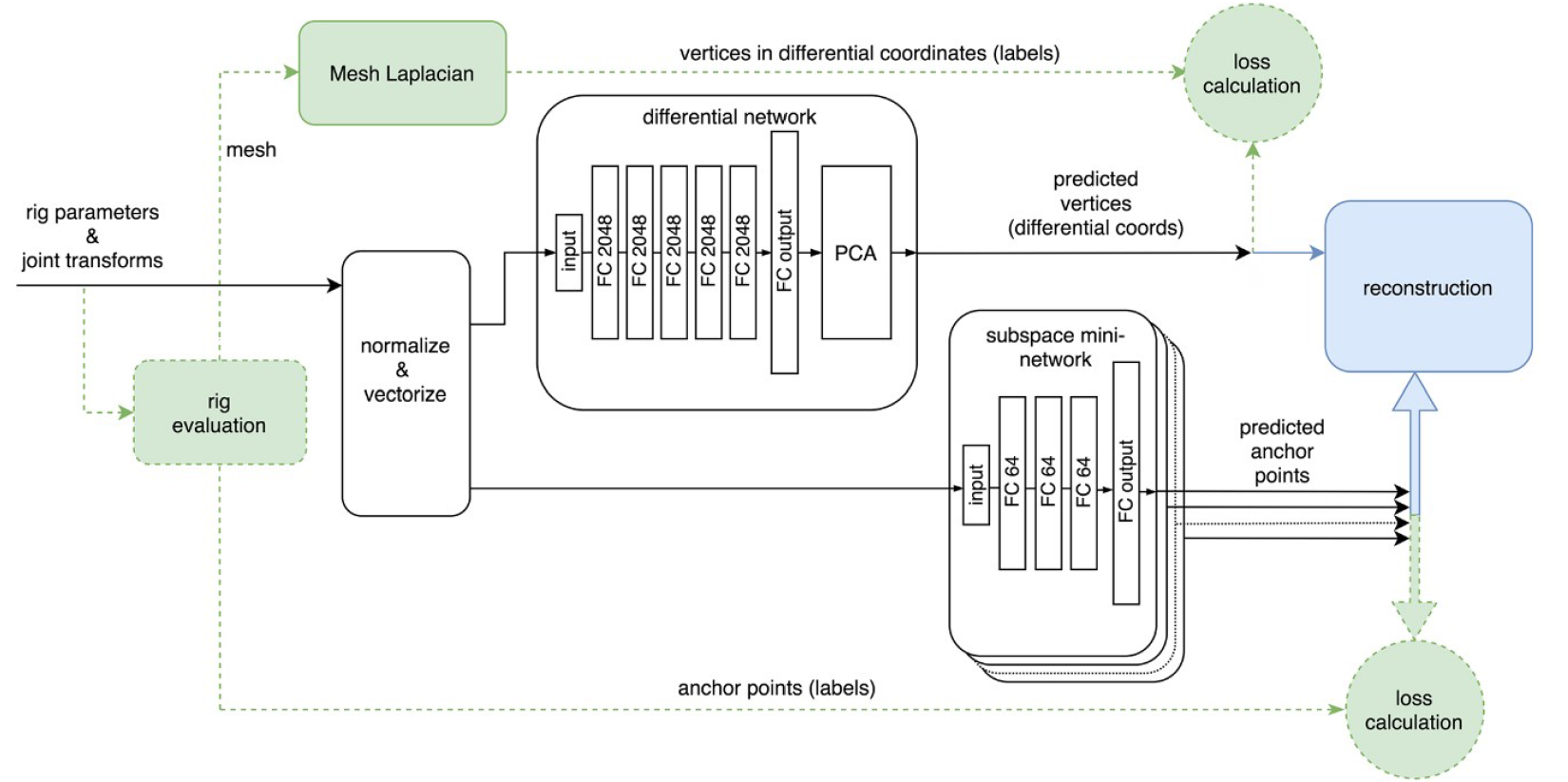}
    \caption{\centering{Training and inference workflow of \cite{bluesky} }}
    \label{fig:networks_bluesky}
\end{figure*}

The same year, another approach using Graph Neural Network (GNN) was presented \cite{graphnet_deformation}. The deformation is separated into linear and non-linear parts, but here the non-linear deformation is modeled by a GNN. Based on a previous SIGGRAPH paper \cite{neuroskinning}
, it aims at learning a generalized deformation system so that it applies easily on unseen character meshes.

Finally, \citet{articulation_neural_blend_shapes} proposed to learn corrective neural blend shapes to improve deformation quality. The shapes are learned using a MeshCNN \cite{meshcnn}, a CNN specifically designed to deal with mesh data. During training, the networks learn the blend shapes as well as how to adjust coefficients to obtain realistic deformations. Interestingly, authors show that this method also generalizes to unseen characters. Inference is also really fast as it only outputs blend shapes coefficients.

\section{Our Second ML Deformer}
\subsection{The Chosen One}
To reduce the development risks, get reliable results and provide long-term support, we focused on a method that required regular NN building blocks directly available in off-the-shelf libraries such as \textit{Torch} or \textit{TensorFlow}. Consequently, we ruled out the MeshCNN method which would require us to extract the part about neural blend shapes generation and to implement the MeshCNN paper itself within \textit{Torch}. Then, to ensure data privacy and prevent copyright infringement, we only wanted methods for which the training relies on the studio data and is done internally. This eliminated the GNN paper as well. Additionally, those two papers did not present any comparison metrics which would help to conclude they perform better or are more accurate.

Comparing the remaining papers, \citet{bluesky} stood out clearly: it can be used to learn both facial and body deformations from both rig controllers and/or skeleton. It also proved to provide more reliable body deformation compared to \citet{dreamworks_body} since it uses one network to model the entire mesh (see Fig.\ref{fig:comparaison_methods_deformation}). Finally, they present comparisons for facial deformation which show that it also performs better than other off-the-shelf methods (see Fig.8 and Table 6 of the paper).

\subsection{A New Hope}

\citet{bluesky} takes rig parameters and bone transforms as input and learns non-linear deformations in Cartesian coordinates. Inputs are then fed to two kinds of network as shown on Fig.\ref{fig:networks_bluesky}. First, a \textit{Differential Network} learns to infer how the mesh moves in differential coordinates. The values of these coordinates are obtained by applying a Laplacian operator over the mesh position in Cartesian coordinates \cite{sorkine_laplacian}. This network is composed of dense layers and a PCA layer. With the differential coordinates, the network learns high frequency deformations over the entire mesh but it can fail at learning low frequency ones. To account for this, subspace mini-networks are used to learn displacement of vertices in Cartesian coordinates and correct the position of the mesh vertices during reconstruction. 

Since this method was developed with a similar objective than \cite{dreamworks_body}, i.e. speed up the rig evaluation, and our first prototype confirmed it, we figured we could use it for an even more impactful purpose than crowds: accelerating rigs evaluation for animators. Rig performance appears to be a common issue, and business wise it is a broader target. We were quite enthusiastic about our first results and thought \textit{"let's create a new product integrated in Maya and target animation departments"}.

\section{Trying to Speed Up Animation with a ML Deformer}

Now that we picked our method and knew our market, we had to implement and package it in a minimum viable product referred as \textit{Golaem Deep} in the following. The product has to allow studios to extract and train the model with their own animations whatever the rig, be accurate enough for animators to use it as an animation proxy and faster enough to justify integrating it in the pipeline (and paying licenses for it). Last but not least, contrary to a crowd scenario, the model has to generalize with unseen animations.

\subsection{Which Framework for Training / Inference?}
\label{sec:soft_ml_library}
A particular attention was paid to pick the most appropriate framework to develop with. Especially since we required a complete C++ workflow to favor computational efficiency as well as code obfuscation. We started evaluating \textit{TensorFlow} and \textit{Torch}, which are the most used Python libraries when it comes to deep learning. While they are quite equivalent in terms of feature set, \textit{TensorFlow} did not provide a C++ API. \textit{Torch} has a C++ API which is labeled "beta" as of today but it contained all the components we needed with a few minor exceptions. We also considered \textit{ONNX} but it could only perform inference. Since we wanted a simple dependency workflow using only one library, we ruled it out.

As mentioned before, all the main elements necessary to implement the NN defined in \cite{bluesky} were present in the \textit{Torch} C++ library as well as the PCA. Yet, we had to implement a Cholesky decomposition of the full-rank Laplacian matrix solver (based on the \textit{Torch} Python implementation). See §3.2.5 of the paper for more details.

\subsection{Extracting the Data and Training the Model}

\begin{figure}
    \centering
    \includegraphics[trim = {0cm 0cm 0cm 0cm}, clip, width = 1\columnwidth]{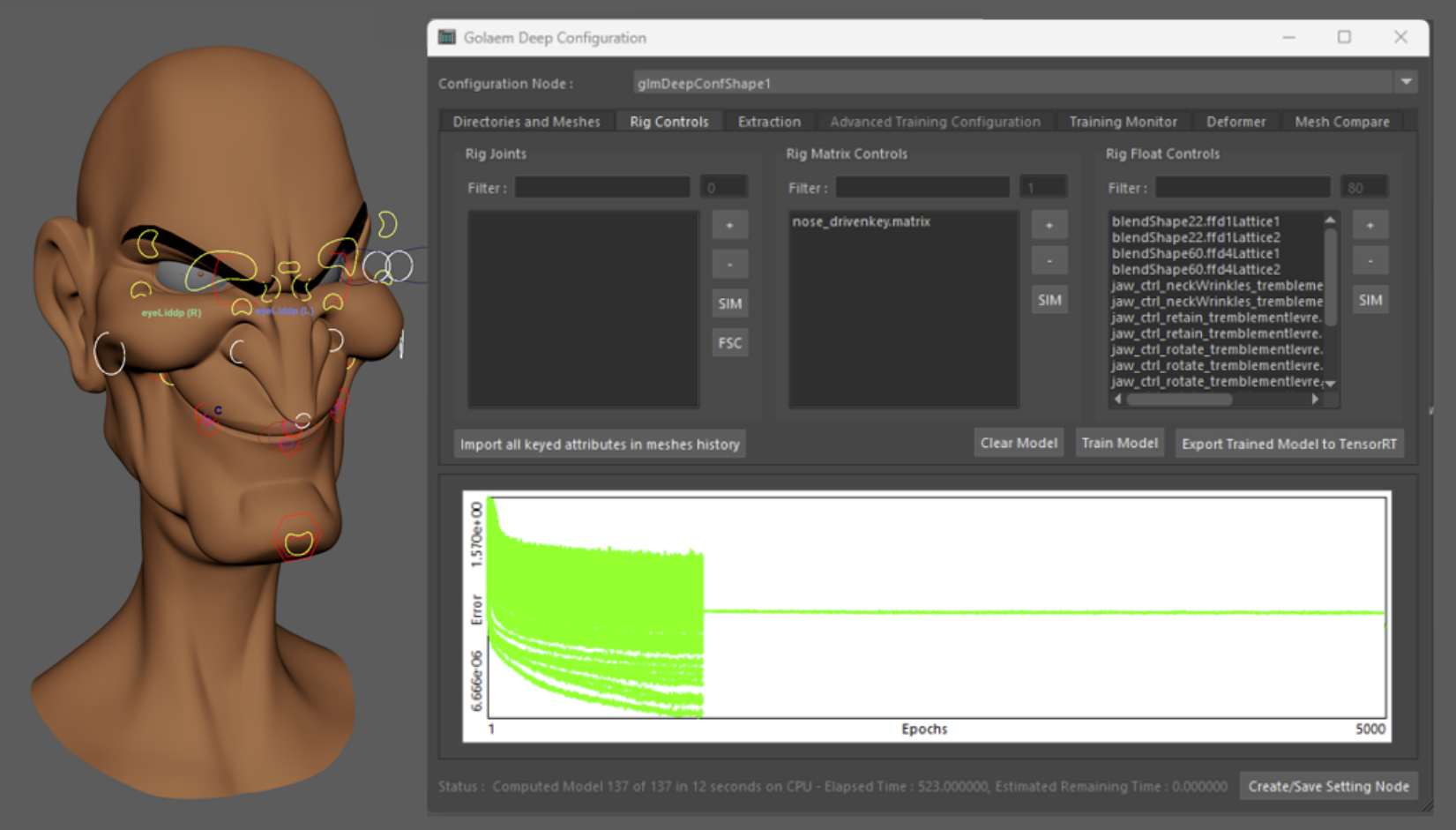}
    \caption{\centering{Example of data extraction and training for a character face in  \textit{Golaem Deep}.}}
    \label{fig:deepUi}
\end{figure}

Extracting the relevant data is crucial when it comes to  NN training and accurate inference. Though, considering there are as many ways of rigging a character than there are riggers (at least), it's no picnic to fit them all! The chosen workflow is the following:
\begin{itemize}
    
    \item first, the user lists all the character meshes that need to be approximated in the output tab. For each of those meshes, a linearly deformed version is built automatically. By default, every output mesh is approximated by a different NN. Yet, for items composed of multiple meshes, the user can specify to use the same one;
    \item then, the user lists all the rig controllers and joint attributes, scalars or matrices, that are contributing to some deformations on the character meshes in the input tab (Fig.\ref{fig:deepUi}). Those attributes can be detected automatically if they are keyframed and connected to one of the output mesh;  
    \item finally, for each animation scene, the input attributes and the vertex positions of the final and linear meshes are extracted into a CSV-like file. The dataset can be extended any time a new scene is extracted within the same directory and the training can always be resumed from the last checkpoint; 
\end{itemize}

Once the data is extracted, training and validation sets are automatically sampled from it (every 1/nth frame). Mental note: as animations are continuous, it would be more relevant to let the user define those in order to validate model generalization. The training is done using \textit{Torch} C++ and \textit{CUDA} acceleration when available. For a character rig composed of 266 inputs, 6840 vertices, and a 136 subspace network, the training takes 60min for 5000 epochs on a \textit{NVIDIA GeForce RTX 4060} GPU. Once trained, the ML deformer is applied to the linear mesh and can be used as an animation proxy.

\subsection{Some Success: Accuracy}

\begin{figure}
    \centering
    \includegraphics[trim = {0cm 0cm 0cm 0cm}, clip, width = 1\columnwidth]{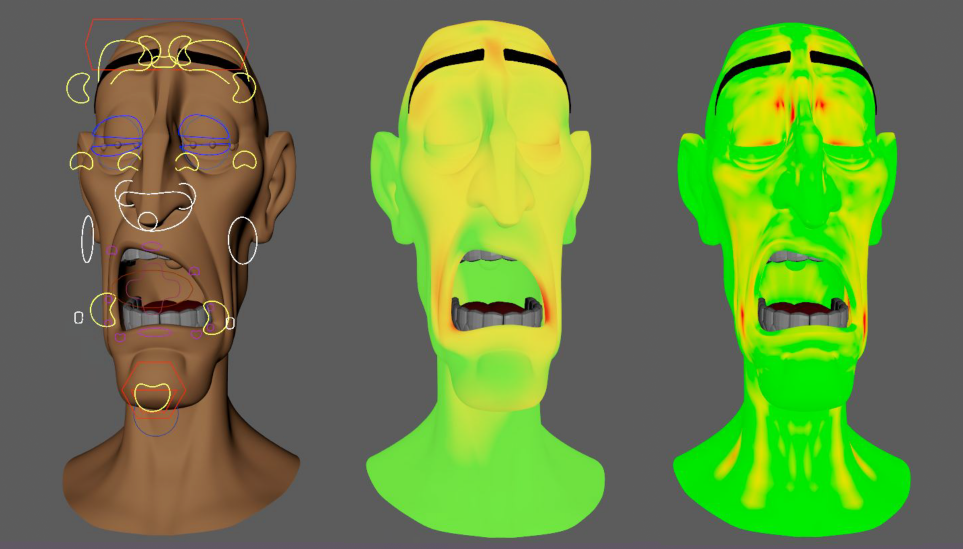}
    \caption{\centering{Target animation (left), ML deformer error heat map (middle), ML deformer uncertainty heat map (right).}}
    \label{fig:edgarresults}
\end{figure}

For the purpose of testing our deformer in production conditions, we partnered with 5 animation studios covering the full range of show types; from top-tier long feature animated movies to preschool TV shows. Using their assets, we could successfully conclude that our method is able to generate accurate deformations and to generalize to previously unseen scenes. Fig.\ref{fig:edgar_faces} shows some results that we obtained for facial deformations: wrinkles on the neck, the forehead or the cheeks are well reproduced for each frame, even for pronounced movements.

Deformation accuracy can be verified using a vertex position error heat map as shown on Fig.\ref{fig:edgarresults}. If large errors are observed, more animation data might be needed to improve the results. In that case, qualifying the missing animation can be challenging so we added a feature to guide the user which displays the \textit{epistemic uncertainty} of the NN through a heat map. This uncertainty describes how well the network learned during training: if the uncertainty is high for a particular zone, more frames or more diverse animations are required involving those deformations. It directly indicates what kind of animation must be added to the training set in order to improve the general quality performance of the method. As an example, Fig.\ref{fig:edgarresults} shows that the inferred deformations for the wrinkles around the mouth are quite uncertain, and large errors are indeed displayed for this part of the face. Adding more animation of this part of the face in the training set would help decrease this uncertainty and improve the generalization of the deformation.

\subsection{...but some Failure: Performances}
\label{sec:time_perf}
Yet, the main motivation to create this new product is to speed up rig evaluation. In their papers, \citet{dreamworks_body} and \citet{bluesky} mention a 5 to 10 deformation evaluation speed up factor but after weeks of experiments, we came to the conclusion that their rigs were optimized for their model and those factors did not include the rig evaluation itself. As a matter of fact, studios use a wide range of rigging techniques and non linear deformations are just a subset of the overall character complexity; the rig evaluation budget is also spent in solving controller positions, evaluating layers of blend shapes, computing inverse kinematics... While the performances with simple rigs were satisfying, we could only reach a x1.2-x2 speed up with production assets. Thus, it became quite clear we had to optimize our inference process beyond what is described in the paper in order to deliver a relevant product.

We tried several things such as optimizing the data exchange between GPU and CPU, using float16 instead of float32 for the network's weights, and using \textit{NVIDIA TensorRT} for inference instead of \textit{Torch}. Tables \ref{tab:time_perf} and \ref{tab:time_optim} present our results with the two production assets we were allowed to disclose. We compare the total time spent in the deformer depending on the hardware used. Column 3 of Table \ref{tab:time_perf} shows the initial fps before any optimization and column 4 shows the fps using our deformer. Tests that have been performed with \textit{Maya 2022}, in parallel evaluation mode, over 5000 frames, with a \textit{NVIDIA GeForce RTX 4060} GPU and 16 CPUs. 

\begin{table}
    \centering
    \begin{tabular}{|c|c|c|c|c|}
        \hline
         & Number of & Number of & \textit{Maya} & With ML  \\
          & inputs & vertices & only & Deformer \\
        \hline
       CharA  & 266 & 6840 & 8.7 fps & 22.1 fps \\
       \hline
       CharB & 1633 & 7186 & 10.1 fps & 18.6 fps \\
       \hline
    \end{tabular}
    \caption{\centering{Evaluation performances on two production assets.}}
    \label{tab:time_perf}
\end{table}

\begin{table}
    \centering
    \begin{tabular}{|c|c|c|c|c|c|c|c|c|}
        \hline
         & \textit{Torch} CPU & \textit{Torch} GPU & \textit{Torch} GPU & \textit{NVIDIA} \\
          &  & (float32) & (float16) & \textit{TensorRT} \\
        \hline
       CharA  & 12.39 ms & 12.49 ms & 12.34 ms & 12.33 ms\\
       \hline
       CharB & 19.33 ms & 15.35 ms & 15.57 ms & 15.55 ms\\
       \hline
    \end{tabular}
    \caption{\centering{Optimization  results on two production assets.}}
    \label{tab:time_optim}
\end{table}

The first takeaway from those results is that none of the implemented optimization presents a significant improvement on the overall rig evaluation. No differences are observed for CharA, and using the GPU slightly improves performances for CharB. Since CharB has a higher number of outputs and much more input values, the differential model is larger than the one for CharA. The computing part due to the differential network is always the limiting factor, so this difference in model size explains the difference in speed gain. On the contrary, if the differential model is small, it is faster to only use the CPU as it discards the CPU/GPU exchange time, which could be longer than the time gained with the GPU to infer the model. The other production assets we tested confirmed those results.

In the end, we found it difficult to assess which setting works best as it depends on a large number of factors varying from one studio to another, such as:
 the deployed Maya version and how the rig complies with Parallel evaluation, the amount of non linear deformations in the rig, the number of input controller values, the number of output mesh vertices, the number of subspace networks, the hardware used for inference (type of GPU and number of CPUs)... Unfortunately, a potential gain of maximum factor 2 in rig evaluation speed is not sufficient to justify a new product. So we went back to plan A.

\section{Back to ML Deformer for CG Crowds}

As stated before, CG crowds deformations are limited to a subset of predefined nodes such as skinning, blend shapes and geometry caches. Thus, integrating our ML deformer in our crowd package will allow to compress those extra deformations and replicate them on the simulated characters.

\begin{figure*}
    \centering
    \includegraphics[trim = {1cm 9cm 1cm 3.5cm}, clip, width = 0.9\textwidth]{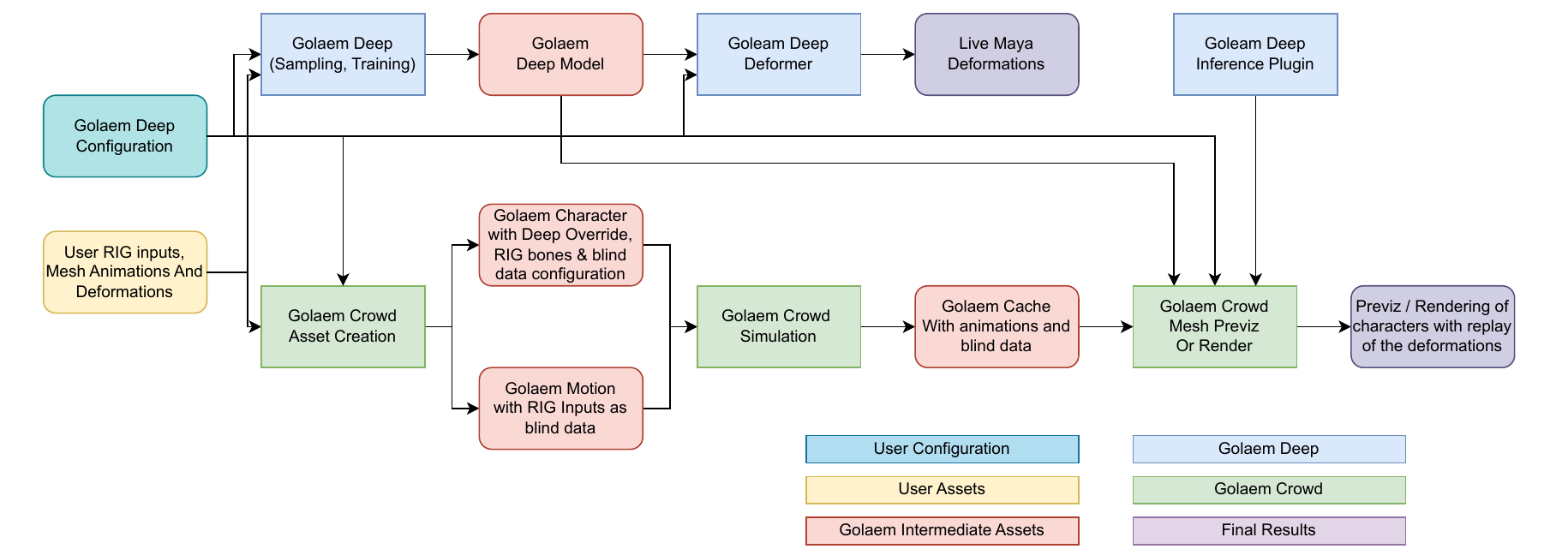}
    \caption{\centering{Workflow of the ML deformer integrated in \textit{Golaem Crowd} and trained with \textit{Golaem Deep}.}}
    \label{fig:DeepInCrowdWorkflow}
\end{figure*}

Even if we decided not to make a product out of  \textit{Golaem Deep}, we rely on it to generate the trained NN model based on the rig inputs and its deformations (see Fig.\ref{fig:DeepInCrowdWorkflow}). On the  \textit{Golaem Crowd} side, we extended our Character Asset file format with a \textit{Deep Override} node which stores the path to a configuration file. This file references which existing meshes will be influenced by the deformer and the name of the rig controller inputs. Then, we extended our Motion Asset file format to also store the rig controller input values for each frame of a clip (in addition to joint and blend shapes animations). 

Like any Character Asset node, the ML deformer can be enabled / disabled per mesh or per character and based on user defined rules, such as camera distance at render time. As our crowd package relies on an \textit{OpenGL} vertex shader for real time character display at simulation/layout time, it is not possible to preview the result of the ML deformer. Those can be seen at render time or if using our render geometry display mode (i.e. less efficient than the \textit{OpenGL} display mode). 
Finally, it is interesting to notice that, in most situations, crowd artists prefer the ML deformer to overfit the data. As the simulation engine is mostly used to blend clips together, it does not really generate unseen animations and does not require the model to generalize. This is not true anymore when the shot takes advantage of physics simulation such as ragdolls for example. Using this workflow, we were able to approximate any production rig within our crowd package (see Fig.\ref{fig:edgar_faces}) but there are still some aspects to be polished and streamlined in the upcoming releases.

\section{Discussion and Future Works}
\subsection{Inference Performance}
\label{sec:discussion_inf_perf}
We spent a large amount of time trying to optimize our inference process but it was definitely not a piece of cake. As explained in section \ref{sec:time_perf} the method takes into account parameters that can largely vary from one rig to another and have a huge influence on the inference time. Trade-offs need to be found between speed and accuracy; in the future, we plan to provide a slider which will automatically set the NN hyper parameters based on those two metrics.

Implementing another method could also be a game changer. It is very likely that \citet{articulation_neural_blend_shapes} performs better than our current approach. The number of neural blend shapes learned during training and output at inference is the most important parameter in this context. Thus, we would only need to predict a certain number of blend shapes weights for each frame, which is computationally lighter than what needs to be predicted currently (number of vertices). In addition, according to the \textit{Third System Effect}, a software can not be successful if it has not been implemented three times\footnote{\url{https://www.johndcook.com/blog/2011/04/18/third-system-effect/}}.

Finally, in our crowd package, inference is currently done sequentially for each character. As crowd characters usually share meshes, we plan to batch the inference to reduce computation time.

\subsection{Training Data}
\label{sec:discussion_training_data}
We voluntarily eluded the question of training data up to this point but it requires a particular discussion.  Since we only use our deformer in our crowd package, users mostly need to overfit the deformations perfectly on their crowd characters. This is why we choose to only support keyframe animation for training data. As of today, we estimate that a minimum of 40s of animation are required to obtain satisfying results. It is obviously highly dependent on the diversity and complexity of the animation sequences used for the training, as well as on the complexity of the rig.

However, to create the training data, the authors recommend to generate random poses, sampling the range of possible values for each rig controller. If the user is not able to provide enough animations or needs to do more than blending clips together (physics simulation, procedural FK/IK/LookAt edits...), it could be useful to provide such a tool. 

\subsection{ML Deformers in Other DCCs}
At the time we started our project (early 2021), no off-the-shelf DCC included a ML deformer. \textit{Epic Games} were the first to release one in UE 5.0 in April 2022, and \textit{SideFX} followed in October 2023 in Houdini 20.0.

The ML deformer of \textit{Epic Games} targets rig portability: it uses a \textit{Maya} plugin to generate the necessary data and does the training and inference within \textit{UE}. Three methods can be used to learn the non-linear deformations: a vertex delta model, a neural morph model and a nearest neighbour algorithm. From our tests, the vertex delta model refers to \citet{bluesky} and the neural morph model is linked to the MeshCNN method presented earlier \cite{articulation_neural_blend_shapes}. These methods can all be used within the \textit{ML Deformer Framework}. Having these different methods allows to choose the most relevant one for each scenario (performance vs accuracy). This workflow has recently been used in the \textit{Fortnite} game for muscle and cloth approximations\footnote{\url{https://www.unrealengine.com/en-US/blog/battle-testing-unreal-engine-5-1-s-new-features-on-fortnite-battle-royale-chapter-4}}. 
\textit{ONNX} Runtime is used for inference and \textit{Torch} Python for training. 

On the other hand, the ML deformer of \textit{Houdini} targets  performances and proposes to bake heavy character simulations in a NN. For now it is available as an experiment of the content library\footnote{\url{https://www.sidefx.com/contentlibrary/ml-deformer/}}. All the workflow fits within \textit{Houdini} and uses internal tools developed to enhance performances such as PCA and \textit{ONNX}-based inference geometry nodes. Similarly to \textit{UE}, they use \textit{Torch} for training and \textit{ONNX} Runtime for inference. From our understanding, they also implemented \citet{bluesky}. 

\section{Conclusion}
Implementing a ML deformer has been a much more challenging task than we anticipated to say the least.
We spent a fair amount of time investigating the different options we had in terms of methods as well as software libraries and learned a lot during this process.  Our main disappointment was to observe that we could not obtain significant speed performance gain using this deformer. This having been said, we will keep trying  and implement a third ML deformer method in the near future.

Deep learning can be intimidating at first sight for users. Thus, in order to make the tool available to non technical artists, we took particular care to design a simple workflow, requiring few to no machine learning skills. We also provided artist-friendly features allowing to check and analyze results when necessary. For example, the quality or the completeness of the training data can be assessed using the \textit{epistemic uncertainty} of the network. This guides the user about what kind of data has to be produced to improve deformation accuracy. Another key feature of the tool is that the training relies on the studio data and is done internally. This ensures data privacy and prevents from copyright infringement. 

In the end, we were able to successfully implement  this ML deformer in \textit{Torch} C++ and include it in the \textit{Golaem Crowd} workflow. It enables very fine and stylized animation on crowd characters like never before. This will allow studios to move the camera much closer to their CG crowds, and create more amazing shots. 

\begin{acks}
We would like to thank \textit{Stim Studio}  who provided us with assets and animations we could use to test, develop and illustrate our tool in this paper. 
\end{acks}

\bibliographystyle{ACM-Reference-Format}
\bibliography{deep_crowd}



\end{document}